\documentclass[preprint]{article}
\usepackage{epsfig} 
\long\def\@makecaption#1#2{ 
  \begin{small} 
  \vskip\abovecaptionskip 
  \sbox\@tempboxa{#1. #2} 
  \ifdim \wd\@tempboxa >\hsize 
     #1. #2\par 
  \else 
    \global \@minipagefalse 
    \hb@xt@\hsize{\hfil\box\@tempboxa\hfil} 
  \fi 
  \vskip\belowcaptionskip 
  \end{small}}

\newlength{\ppp} 
\newcommand{\hp}{\hspace*{\ppp}} 
\begin{document} 
\thispagestyle{empty} 
 
\twocolumn[ 
\begin{small}
\hspace{6cm}
{\it 8th International Conference on Quasicrystals, Bangalor sept. 2002}

\hspace{6cm}
{\it submitted to Journal of  Non-Crystalline Solids}
\end{small}

\vspace*{10mm} 
\begin{LARGE} 
%\begin{center} 
% 
%  TITLE 
% 
{\bf Indirect Mn-Mn pair interaction induces pseudogap  
in Al(Si)-Mn approximants} 
% 
%  END TITLE 
% 
\vspace{0.6cm} 
%\end{center} 
\end{LARGE} 
 
\begin{large} 
%\begin{center} 
% 
%  AUTHORS 
% 
Guy Trambly de Laissardi\`ere$^{a}$
\footnote{},
Duc Nguyen-Manh$^{b}$,  
Didier Mayou$^{c}$ 
% 
%  END AUTHORS 
% 
 
%\end{center} 
\end{large} 
\begin{footnotesize} 
\begin{it} 
%\begin{center} 
% 
%  ADDRESS 
% 
\vspace{0.15 cm} 
a) Laboratoire de Physique Th\'eorique et Mod\'elisation, 
CNRS-Universit\'e de Cergy-Pontoise
(UMR 8089), Neuville, 95031 Cergy-Pontoise, 
France.\\  
b) Department of Materials, University of Oxford, Oxford,  
Parks Road, OX1 3PH, United Kindom.\\ 
c) Laboratoire d'Etudes des Propri\'et\'es Electronique 
des Solides (CNRS), 38042 Grenoble C\'edex 9, France.\\ 
% 
%  END ADDRESS 
% 
 
%\end{center} 
\end{it} 
\end{footnotesize} 
\begin{footnotesize} 
%\begin{center} 
% 
%  DATE 
% 
2002/9/6 %received in revised form 
% 
%  END DATE 
% 
 
%\end{center} 
\end{footnotesize} 
\vspace{4ex} 
\begin{small} 
%\hrule\vspace{3ex} 
\begin{minipage}{\textwidth} 
%{\bf Abstract}\vspace{2ex}\\ 
\hp 
% 
%  ABSTRACT 
% 
 
The effect on the electronic structure  
of an indirect Mn-Mn interaction mediated by  
the valence states and the sp-d hybridisation 
is presented. In Al(rich)-Mn phases related  
to quasicrystals ($\rm Al_{12}Mn$, o-$\rm Al_6Mn$, 
$\alpha$-$\rm Al_9Mn_2Si$),  
this indirect interaction creates a Hume-Rothery 
pseudogap in the density of states together with 
a minimisation of the band energy. 
It is shown that the Mn-Mn interaction  
up to the distance around  
10-20\,$\rm \AA$ plays
an essential role in stabilizing 
related quasicrystal
structures.

% 
%  END ABSTRACT 
% 
\vspace{2.5ex} 
\end{minipage}\vspace{3ex} 
%\hrule 
\end{small}\vspace{6ex} 

] 
\footnotetext{
{\it Correspondence to:}
Dr. G. Trambly de Laissardi\`ere:
Laboratoire de Physique Th\'eorique et Mod\'elisation,
Universit\'e de Cergy-Pontoise, Neuville III-1,
5 Mail Gay-Lussac,
95031 Cergy-Pontoise, France.\\
Tel: 33 1 34 25 69 65\\
Fax: 33 1 34 25 70 04\\
guy.trambly@ptm.u-cergy.fr}

% 
%  MAIN TEXT 
% 

\noindent  
{\bf 1. Introduction} 
\vspace{0.6cm} 

Al(rich)-Mn  and Al(rich)-Si-Mn systems,  
contain many crystalline approximants of  quasicrystals.  
These phases are good examples to analyse the  
effect of the position of transition metal (TM) 
atoms in stabilizing complex structure  
related to quasiperiodicity. 
The origin of the stabilization of quasicrystals is  
still unclear in spite of many experimental and theoretical 
study.  
For Al-based quasicrystals,  
a Hume-Rothery mechanism   
\cite{Massalski78,Paxton97}  have been shown to play a significant  
role (see for instance 
\cite{Mayou94,Belin02,Mizutani02} and Refs. within).  
In these phases, the average number of electron per 
atom (ratio $e/a$) is an important parameter.  
Indeed, the occurrence of phases related to quasicrystals is  
explained by the fact that they are  
electron compounds with similar $e/a$ ratio in spite 
of different constituents and different atomic concentrations  
\cite{Tsai91,Gratias93}.  
A band energy minimisation occurs when the Fermi sphere 
touches a pseudo-Brillouin zone, constructed by Bragg  
vectors ${\bf K}_p$ corresponding to intense peaks  
in the experimental diffraction pattern. 
The Hume-Rothery condition for alloying  
is then $2k_F \simeq K_p$. Assuming a free electron  
valence band, the Fermi momentum, $k_F$, is calculated  
from $e/a$. 
 
In sp Hume-Rothery alloys,  
valence electrons (sp electrons)   
are nearly free. 
Their density of states (DOS)  is well 
described by the  Jones theory 
\cite{Massalski78,Paxton97}. 
The Fermi-sphere\,/\,pseudo-Brillouin zone 
interaction creates a depletion in the DOS, 
called pseudogap, 
near the Fermi energy $E_F$. 
Such a pseudogap has been found experimentally  
and 
from first-principles calculations in 
many sp quasicrystals and approximants 
\cite{Fujiwara91,Hafner92,Mizutani02}. 
It has also been found in many  
icosahedral approximants  
containing TM elements 
\cite{FujiAlMnSi,Krajci97_dAlPdMn,Mizutani02} 
whereas there are contradictory 
results  about decagonal  
phases (Ref. \cite{Guy_beta} and
Refs. within). 
But, the treatment of 
Al(rich)-TM is more complicated as 
d states of TM are not nearly-free states. 
In the case of crystals and icosahedral quasicrystals 
it has been shown \cite{GuyEuroPhys93,GuyPRB95} 
that sp-d hybridisation  
increases a pseudogap.  
In some particular  
cases a pseudogap may also be induced by the sp-d  
hybridisation \cite{Duc92,HafnerICQ8}. 
 
The Hume-Rothery stabilization 
can also be viewed as a 
consequence of oscillating pair interactions between 
atoms (Refs. \cite{Mayou94,Galher02} and Refs. in there).  
In this direction  
Zou and Carlsson have shown that an indirect Mn-Mn  
interaction, mediated by sp states of Al,  
is strong enough to favours  Mn-Mn  
distances close to 4.7\,$\rm \AA$ in Al(rich)-Mn 
quasicrystals and approximants. 
Here, it is shown that an indirect 
Mn-Mn interaction up to 10-20\,$\rm \AA$  
induces pseudogap at $E_F$ in the approximants:  
cubic $\rm Al_{12}Mn$ \cite{GuyPRB95},  
orthorhombic o-$\rm Al_6Mn$ \cite{GuyPRB95}, 
and cubic $\alpha$-$\rm Al_9Mn_2Si$  
\cite{Cooper66}. 
The importance of Mn-Mn interaction up to large distances 
shows the complexity of the stabilizing process.  
Obviously ``frustration'' mechanism should occur that 
may favour for complex atomic structures. 
As $\rm Al_{12}Mn$, o-$\rm Al_6Mn$ and  
$\alpha$-$\rm Al_9Mn_2Si$ are related to quasicrystals, 
this study suggests that 
a Hume-Rothery stabilization, 
expressed in terms of Mn-Mn interaction, 
is intrinsically  
linked to the emergence of quasiperiodic structures 
in Al(Si)-Mn systems.

\vspace{1cm} 
\noindent 
{\bf 2. Effective Bragg potential for Al(rich)-Mn alloys} 
\vspace{0.6cm} 
 
For sp Hume-Rothery alloys, the valence states (sp states) 
are nearly-free states scattered by a weak potential  
(Bragg potential, $V_B$). 
In this section, we show briefly that in sp-d Hume-Rothery  
alloys, sp electrons feel an  
{\it ``effective Bragg potential''}  
\cite{GuyPRB95,Guy_beta} 
that takes into account the strong effect TM atoms  
via the sp-d hybridisation. 
 
Following a classical approximation 
\cite{Friedel56,Anderson61}  
for Al(Si)-Mn alloys, a simplified  
model is considered where sp states are nearly-free  
and d states are localized on Mn sites $i$. 
The effective hamiltonian  
for the sp states is written: 
\begin{eqnarray} 
H_{eff(sp)}= \frac{\hbar^2\,k^2}{2m} + V_{B,eff} 
\label{Hamil_eff_sp} 
\end{eqnarray} 
where $V_{B,eff}$ is an effective Bragg potential  
that takes into account the scattering  
of  sp states by the  
strong potential ot Mn atoms.  
$V_{B,eff}$ depends thus  on  
the positions ${\bf r}_i$ of Mn atoms. 
Assuming that all Mn atoms are equivalent and that two Mn  
atoms are not first-neighbour, one obtains  
\cite{GuyPRB95,Guy_beta}: 
\begin{eqnarray} 
{ V_{B,eff}({\bf r}) =  \sum_{\bf K}  
V_{B,eff}({\bf K}) e^{i{\bf K}.{\bf r}}, } \\ 
{ V_{B,eff}({\bf K})  =   
V_B({\bf K}) +  
\frac{|t_{{\bf K}}|^2}{E - E_d}  
\sum_i e^{-i {\bf K}.{\bf r}_{i}}, } 
\label{EqVeffectif} 
\end{eqnarray} 
where the vectors ${\bf K}$ belong to the  
reciprocal lattice, $t_{{\bf K}}$ is a average 
matrix element that  
couples sp states ${\bf k}$ and ${\bf k}-{\bf K}$ via the 
sp-d hybridisation, and $E_d$ is the energy of d states. 
The term $V_B({\bf K})$ is a weak potential independent with the  
energy $E$. It corresponds to the  Bragg potential for 
sp Hume-Rothery compounds. 
 
The last term in equation (\ref{EqVeffectif}), is due to  
the d resonance of the wave function by the potential of  
Mn atoms. It is strong in 
an energy range 
$ E_d-\Gamma \leq E \leq E_d+\Gamma$, 
where 
$2\Gamma$ is the width of the d resonance. 
This term is essential as  
it does represent 
the diffraction of the sp electrons by a network 
of d orbitals, 
i.e. the  factor 
$\left(\sum_i e^{-i {\bf K}.{\bf r}_{i}}\right)$ 
corresponding to the structure factor of the 
TM atoms sub-lattice. 
As the d band of Mn is almost half filled, 
$E_F \simeq E_d$, this factor is important 
for energy close to $E_F$. 
Note that  
the Bragg planes associated with the second term of 
equation (\ref{EqVeffectif}) correspond to 
Bragg planes determined by diffraction. 
 
This analysis shows that both sp-d hybridisation and  
diffraction 
of sp states by the sub-lattice of Mn atoms are essential  
to understand the electronic structure of  
Al(Si)-Mn alloys \cite{Guy_beta}. 
The strong effect of sp-d hybridisation 
on the pseudogap is then  understood   
in the framework of  Hume-Rothery mechanism. 
 
\vspace{1cm} 
\noindent 
{\bf 3. Two Mn in the Al(Si) matrix} 
\vspace{0.6cm} 
 
As a Hume-Rothery stabilization is a consequence of oscillation of  
the charge density of the valence electrons with energy close to 
$E_F$, 
a most stable atomic structure is obtained when distances between  
atoms are multiples of the wavelength of electrons with energy close 
to $E_F$. Since the scattering of valence sp states by the 
Mn sub-lattice is strong, 
the Friedel oscillations of charge of sp electrons 
around Mn  must have a strong effect on a stabilization. 
Therefore a Hume-Rothery mechanism in Al(rich)-Mn compounds 
might be analysed in term of 
a Mn-Mn pair interaction 
resulting from a strong sp-d hybridisation. 
Zou and Carlsson \cite{ZouPRL93,Zou94} have  
calculated this interaction 
from an Anderson 
model hamiltonian with two impurities, using a Green's function 
method. It is found that a specific Mn-Mn distance of  
4.7\,$\rm \AA$ favours for a stabilization of Al-Mn approximants 
\cite{ZouPRL93}. 
As 4.7\,$\rm \AA$ is larger than first neighbour distances, 
this shows the existence of an indirect medium range Mn-Mn  
interaction. 
The indirect interaction is mediated by sp-d  
hybridisation where sp  states are mainly Al states. 
 
\begin{figure} 
\begin{center} 
\includegraphics[width=7cm]{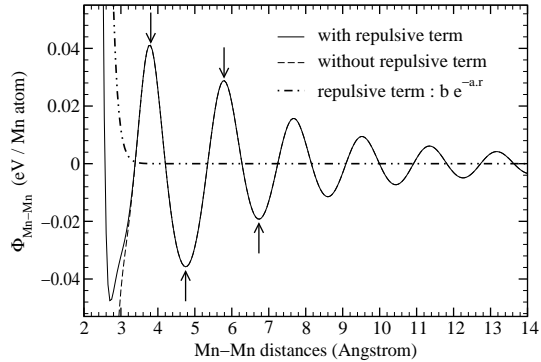} 
%\vspace{3cm} 
\caption{Mn-Mn pair interaction of two non-magnetic manganese atoms in 
free electron matrix, simulating aluminium (and silicon) host. 
$2\Gamma = 2.7$\,eV,  $E_d=11.37$\,eV 
and $E_F=11.7$\,eV.} 
%\vspace{3cm} 
\label{PotMn_Mn_m0} 
\end{center} 
\end{figure}

We calculated 
the indirect Mn-Mn pair interaction $\Phi_{Mn\textrm{-}Mn}$  
from the transfer matrix $T$ of two Mn atoms in the free electrons 
matrix by using the Lloyd formula \cite{GuyPRB97}  
(Fig.\,\ref{PotMn_Mn_m0}). 
According to  classical approximation for metal,  
a phenomenological short range  
repulsive term should be add. 
But this term is not important in the present  
study as we analyse only the medium range order, 
i.e. distances 
larger than first-neighbour distances 
(see Fig.\,\ref{PotMn_Mn_m0}).  
Parameters of the calculation are: 
the Fermi energy $E_F$ fixed by the Al matrix 
($E_F=11.7$\,eV), 
the width of the d resonance $2\Gamma$  
which increases as the sp-d hybridisation increases 
($2\Gamma = 2.7$\,eV), 
and the energy $E_d$ of the d resonance  
which depends on the nature of the transition 
metal atom ($E_d=11.37$\,eV  corresponding to 
$\sim$\,5.8 d electrons per Mn atom). 
A small variation of these parameters does  
not modify qualitatively the results presented 
in the following. 
In this paper only non-magnetic Mn are considered
as most of Mn are non-magnetic in quasicrystals and  
approximants \cite{Virginie2,Hippert99,Prejean02}. 
In particular 
$\rm Al_{12}Mn$, o-$\rm Al_6Mn$ 
and $\alpha$-$\rm Al_9Mn_2Si$ are 
non magnetic \cite{Virginie2}. 
Because of the sharp Fermi surface 
of Al, $\Phi_{TM\textrm{-}TM}$ oscillated  
(Friedel oscillations of the charge density). 
It asymptotic form 
at large TM-TM distance ($r$) is of the form: 
\begin{eqnarray} 
\Phi_{TM\textrm{-}TM}(r) 
\propto \frac{\cos (2k_F\,r-\delta)}{r^3}\;. 
\label{EqVeffOscil} 
\end{eqnarray} 
The phase shift $\delta$  
depends on the nature of the TM atom and varies from $2\pi$ 
to $0$ as the d band fills.  
Magnitude of the medium range interaction is larger  
for Mn-Mn than 
for other transition metal (Cr, Fe, Co, Ni, 
Cu), because the number of  d electrons  
close to $E_F$ is the largest for Mn,  
and 
the most delocalized electrons are electrons with 
Fermi energy.  
The effect analysed here 
is then more important for Al-rich alloys containing  
Mn element than for alloys containing other 
TM elements. 
 
\begin{figure} 
\begin{center} 
\includegraphics[width=7cm]{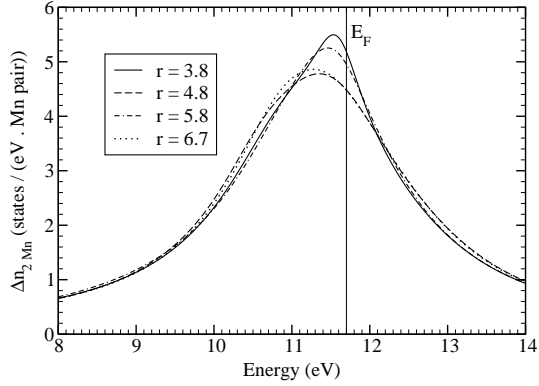} 
%\vspace{3cm} 
\caption{Variation of the DOS,  
$\Delta n_{2Mn}(E)$, due to 2  
Mn impurities in free electrons matrix. 
The Mn-Mn distances $r = 3.8\,$${\rm \AA}$ and 
$ r = 5.8$$\rm \,\AA$ correspond to 
positive Mn-Mn interaction, 
whereas $r = 4.8$$\rm \,\AA$ and 
$r = 6.7$$\rm \,\AA$ correspond to minima of the interaction 
(see arrows on Fig. \ref{PotMn_Mn_m0}). 
} 
%\vspace{3cm} 
\label{DOS_2Mn_imp} 
\end{center} 
\end{figure}

The total DOS of  
two Mn atoms in the free electrons  
matrix is: 
\begin{eqnarray} 
n(E,r)=n_{sp}^0(E) + \Delta n_{2Mn}(E,r)\,, 
\label{Eq_DOS_2Mn} 
\end{eqnarray} 
where 
$n_{sp}^0$ is the free electron DOS 
and $\Delta n_{2Mn}$, 
the variation of the total DOS due to two Mn atoms. 
$\Delta n_{2Mn}$ depends on the Mn-Mn distance $r$. 
When $r$ is very large (almost infinity),
each Mn are similar to Mn impurity thus:
$\Delta n_{2Mn} = 2 \Delta n_{1Mn}$,
where $\Delta n_{1Mn}$ is the well known Lorentzian  
of the virtual-bound states. 
But small deviation from the Lorentzian occurs 
for finite  $r$. 
On Fig.\,\ref{DOS_2Mn_imp},   
$\Delta n_{2Mn}(E)$ is drawn for different values of  
$r$.  
$r = 3.8\,$${\rm \AA}$ and 
$ r = 5.8$$\rm \,\AA$ correspond to 
positive values of  Mn-Mn interaction  
(Fig.\,\ref{PotMn_Mn_m0}). These distances  
are thus unstable and the corresponding  
DOSs at $E_F$ increase with respect to  
Lorentzian value. 
On the other hand,  
$r = 4.8$$\rm \,\AA$ and 
$r = 6.7$$\rm \,\AA$ are more stable 
(minima of interaction), and the 
corresponding DOSs at $E_F$ are  
lower than the  Lorentzian value.

\vspace{1cm} 
\noindent 
{\bf 4. Effect of Mn sub-lattice on electronic 
structure of approximants} 
\vspace{0.6cm}

%\vspace{1.5cm} 
\noindent 
{\it 4.1. Density of states} 
\vspace{0.6cm} 
 
\begin{figure}
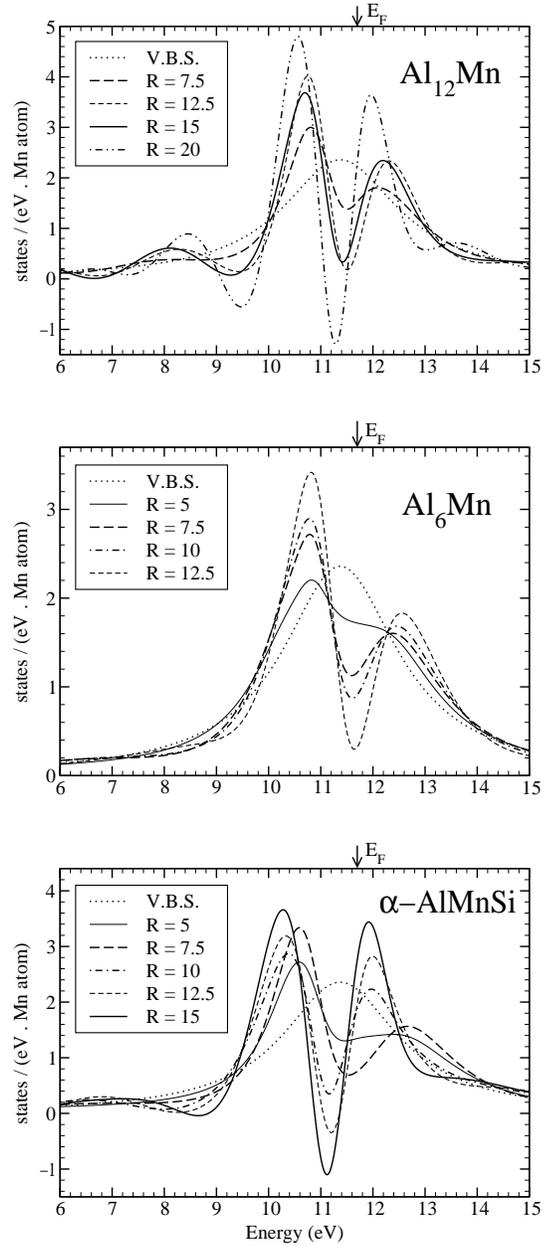
 
%\begin{center} 
\psfig{file=DOS_Al12Mn.eps,width=7cm} 
\vspace{0.2cm} 
 
\psfig{file=DOS_Al6Mn.eps,width=7cm} 
\vspace{0.2cm} 
 
\psfig{file=DOS_alpha.eps,width=7cm} 
%\vspace{1cm} 
\caption{Variation of the DOS  
due to Mn sub-lattice, $\Delta n_{R}(E)$,  
in $\rm Al_{12}Mn$, o-$\rm Al_6Mn$ and  
$\rm \alpha$-$\rm Al_9Mn_2Si$. 
These calculations 
include  effects of  all Mn-Mn pairs up to 
Mn-Mn distance equal to $R$  
($R$ in $\rm \AA$). 
V.B.S. is the Lorentzian of one Mn impurity in 
the free electron matrix (virtual-bound state).} 
\label{DOS_Al12Mn_Al6Mn_alpha} 
%\end{center} 
\end{figure} 
 
In this section, 
the effect of indirect Mn-Mn interaction on the DOS of 
approximants is analysed. 
We focus on the case of cubic $\rm Al_{12}Mn$, orthorhombic  
o-$\rm Al_6Mn$ and 
cubi $\alpha$-$\rm Al_9Mn_2Si$.  
In each of these phases, Mn sites are 
similar and Mn atoms are not 
first-neighbour. 
In metallic alloys, the main aspects of the DOS are consequences of 
short range and  medium range atomic order. 
The effect of the medium range order on the 
pseudogap at Fermi energy is estimated from a simple model  
that takes only into account the Mn-Mn pair effects with 
Mn-Mn distances larger than first-neighbour distances. 
An important question is to determine the distance  
up to which an indirect Mn-Mn 
interaction is essential. 
 
Assuming a Hume-Rothery mechanism for the stabilization, 
the electronic 
energy is a sum of pair interaction. 
As interaction  
magnitudes  
are larger for Mn-Mn than 
for Al-Mn and Al-Al \cite{Mihalkovic96},  
$\Phi_{Mn\textrm{-}Mn}$ 
has a major effect on the electronic 
energy 
and $\Phi_{Al\textrm{-}Al}$,  
$\Phi_{Al\textrm{-}Mn}$ are neglected. 
Triplet effects, quadruplet effects (...), that might be 
important for a transition metal concentration larger than 25\% 
\cite{Widom98}, are neglected. 
In this model, the total DOS, $n_R(E)$,  
is calculated as the sum of the 
variation of the DOS due to each 
Mn-Mn pair:  
\begin{eqnarray} 
\lefteqn{n_{R}(E) =  n_{sp}^0(E) +  \Delta n_R(E),}     \\ 
\lefteqn{\Delta n_R(E) = x \Delta n_{1Mn}(E) }\nonumber \\ 
& & + \sum_{r_{ij} < R}\Big( \Delta n_{2Mn}(E,r_{ij}) - 
2 \Delta n_{1Mn}(E) \Big), 
\end{eqnarray}  
where $i$, $j$ are index of Mn atom, 
$r_{ij}$ is  $\rm Mn_i$-$\rm Mn_j$ distance,
and $x$, the 
number of Mn atoms.
$\Delta n_{2Mn}$ is defined  
by equation (\ref{Eq_DOS_2Mn}). 
$\Delta n_{1Mn}$ is the variation of the DOS due to one 
Mn impurity in the free electron
matrix: virtual-bound state (V.B.S.). 
$\Delta n_{1Mn}$ is a Lorentzian centred at energy $E_d$ 
with a width at half maximum equal to $2\Gamma$. 
$n_{R}$ is the total DOS computed 
by taking into account all  Mn-Mn 
interaction up to Mn-Mn distance equal to $R$.  
$\Delta n_R$ is the part of $n_{R}$ due to Mn atoms. 
 
$\Delta n_{R}(E)$ of  $\rm Al_{12}Mn$,  o-$\rm Al_6Mn$ and  
$\rm \alpha$-$\rm Al_9Mn_2Si$ 
are shown in  
Fig.\,\ref{DOS_Al12Mn_Al6Mn_alpha} for different values  
of distance $R$. 
First Mn-Mn distance is  
6.47\,$\rm \AA$ in $\rm Al_{12}Mn$,  
4.47\,$\rm \AA$ in o-$\rm Al_6Mn$ 
and 4.61 $\rm \AA$ in $\alpha$-$\rm Al_9Mn_2Si$,  
but a well pronounced pseudogap appeared 
only when the Mn-Mn interactions up to  
10-20\,$\rm \AA$ are taken into account. 

Negative  value of $\Delta n_{R}(E)$
induces reduction of the total DOS with 
respect to the free electron value $n_{sp}^0$.
For o-$\rm Al_6Mn$, the minimum of the 
pseudogap corresponds to $\Delta n_{R} \simeq 0$.
The total DOS at the minimum of the pseudogap is 
thus similar to pure Al DOS,
in agreement with first-principles calculation 
\cite{GuyPRB95}. But for $\rm Al_{12}Mn$
and $\alpha$-$\rm Al_9Mn_2Si$,
as $\Delta n_{R} < 0$, a reduction of 
the total DOS with respect to free electron 
case is due to Mn-Mn medium range interaction. 
First-principles studies
\cite{GuyPRB95,FujiAlMnSi} have already 
shown a reduction. 
The present work enlightens a
particular effect of Mn atoms in 
these {\it ab initio} results.

\vspace{0.6cm} 
\noindent 
{\it 4.2. Energy} 
\vspace{0.6cm} 
 
\begin{figure} 
\begin{center} 
\psfig{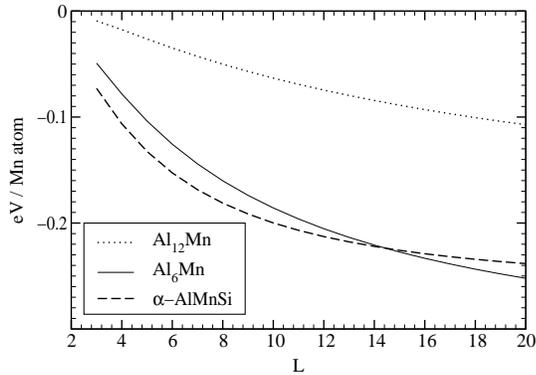} 
%\vspace{1cm} 
\caption{Structural energy ${\cal E}(L)$ of Mn sub-lattice
versus the mean-free path $L$ 
($L$ in $\rm \AA$).} 
\label{FigE_Al12Mn_Al6Mn_alpha} 
\end{center} 
\end{figure}

%As interaction  
%magnitudes  
%are larger for Mn-Mn than 
%for Al-Mn and Al-Al \cite{Mihalkovic96},  
%$\Phi_{Mn\textrm{-}Mn}$ 
%has a major effect on the electronic 
%energy. 
The {\it ``structural energy''}, $\cal{E}$, of the  
Mn sub-lattice in Al host is defined as  
the energy needed to built the 
Mn sub-lattice in the metallic host that simulates  
Al (and Si) host 
from isolated Mn atoms in the same metallic host. 
$\cal{E}$ per unit cell is: 
\begin{eqnarray} 
{\cal E} =  \sum_{i,j\,(j\neq j)}  
\frac{1}{2} \, 
\Phi_{Mn\textrm{-}Mn}(r_{ij})~e^{-\frac{r_{ij}}{L}}\;, 
%{\cal E}_{TM(i)}\;, 
\label{EquationEStruturale} 
\end{eqnarray} 
%where $i$ and $j$ are index of Mn atom and 
%$r_{ij}$,  Mn(i)-Mn(j) distances. 
$L$ is the mean-free path of electrons due to scattering 
by static disorder or phonons \cite{Guy_beta}. 
$L$  depends on the structural quality 
and temperature and can be estimated to be larger 
than 10\,$\rm \AA$. 
${\cal E}(L)$ for  
$\rm Al_{12}Mn$, o-$\rm Al_6Mn$  
and $\alpha$-$\rm Al_9Mn_2Si$ 
are shown on  
Fig.\,\ref{FigE_Al12Mn_Al6Mn_alpha}. 
${\cal E}$ are always negative 
with 
magnitude 
strong enough to give a significant contribution 
to the band energy. 
This result is in good agreement  
with effect of 
Mn-Mn interactions on the pseudogap as shown  
previously. 
According to a Hume-Rothery mechanism, one expects that  
a pseudogap is well pronounced 
for a large value of $|{\cal E}|$.

\vspace{1cm} 
\noindent 
{\bf 5. Conclusion} 
\vspace{0.6cm}

A simple model is presented that  
allows to enlighten  effects of Mn atoms  
on the electronic structure in Al(rich)-Mn phases 
related to quasicrystals.  
It is shown that an indirect Mn-Mn interaction  
up to distances 
10-20\,$\rm \AA$ 
is essential in stabilizing, 
as it creates a Hume-Rothery pseudogap  close to 
$E_F$. 
The band energy is then minimised.  
 
The effect of an  
indirect Mn-Mn interactions has been also study 
in previous works  
\cite{ZouPRL93,Zou94,Mihalkovic96,Guy_beta,GuyPRL00}. 
Recently \cite{Guy_beta},
it explained the origin of a large  
vacancies in the hexagonal $\beta$-$\rm Al_9Mn_3Si$ and  
$\varphi$-$\rm Al_{10}Mn_3$ phases, whereas  
similar site are occupied by Mn in  
$\rm \mu\,Al_{4.12}Mn$ 
and $\rm \lambda\,Al_4Mn$, and by Co 
in $\rm Al_5Co_2$. 
On the other hand, medium range indirect Mn-Mn  
interaction is also determinant for the  
existence or not of magnetic moments in Al-Mn 
quasicrystals and approximants \cite{GuyPRL00}. 
 
As Al(rich)-Mn phase structure are related to  
those of quasicrystals, it suggests  
that a Hume-Rothery stabilization, governs 
by this Mn-Mn interaction, is intrinsically linked 
to the emergence of quasiperiodicity.

%  END MAIN TEXT  
% 
 
\hp 
 
\begin{footnotesize} 
\begin{frenchspacing} 
 
\end{frenchspacing} 
\end{footnotesize} 
 
\end{document}